\documentclass[twocolumn,showpacs,preprintnumbers,amsmath,amssymb,aps,prb]{revtex4}

\usepackage{graphicx}
\usepackage{dcolumn}
\usepackage{bm}

\begin{document}

\title{Transport in chaotic quantum dots: \\ effects of spatial 
symmetries which interchange the leads}

\author{Victor A. Gopar$^{1,2}$, Stefan Rotter$^3$, and Henning
Schomerus$^{1,4}$}
   \affiliation{
$^1$Max-Planck-Institut f\"ur Physik komplexer Systeme,
N\"othnitzer Strasse 38, 01187 Dresden, Germany\\ $^2$Instituto de
Biocomputaci\'on y F{\'i}sica de Sistemas Complejos, Universidad de Zaragoza, 
   Corona de Arag\'on 42, 50009 Zaragoza, Spain\\
$^3$Institute for Theoretical Physics, Vienna University of
Technology, 1040 Vienna, Austria\\
$^4$Department of Physics, Lancaster University, Lancaster LA1
4YB, UK}


\begin{abstract}

We investigate the effect of spatial symmetries on phase coherent
electronic transport through chaotic quantum dots. For systems which have a spatial symmetry
that interchanges the source and drain leads, we find in the framework
of random matrix theory that the
density of the transmission eigenvalues is independent of the
number of channels $N$ in the leads. As a consequence, the weak
localization correction to the conductance vanishes in these
systems, and the shot noise suppression factor $F$ is independent
of $N$. We confirm this prediction 
by means of numerical calculations for stadium
billiards with various lead geometries. These calculations also uncover 
transport signatures of partially preserved symmetries.

\end{abstract}


\maketitle

\section{\label{intro}Introduction}

\begin{figure}[!t]
\begin{center}
\includegraphics[width=\columnwidth]{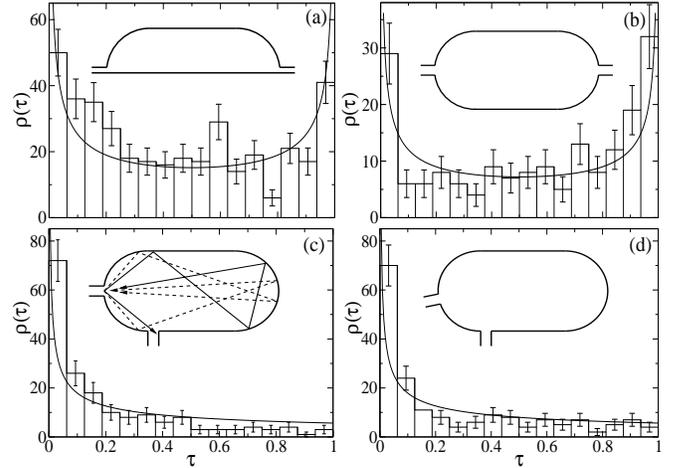}
\end{center}
\caption{\label{fig:systems} 
Distribution of transmission eigenvalues $\rho(\tau)$ (normalized to the
sample size), $N=1$, in stadium geometries
with injection from the left: lead-transposing left-right symmetry (a, b)
and without global symmetry (c, d). The numerical data (histogram) is
compared with the corresponding RMT-predictions (solid lines). Due to the
horizontal alignment of the entrance lead in (c) a fraction of back-reflected
paths have an up-down reflected partner trajectory of equal length (see, 
e.g.,~pair of dashed lines). Although for a class of trajectories the corresponding
partner is cut short by the exit lead (see, e.g., pair of solid lines) 
transport is
strongly influenced by this partial symmetry (see text). Cavity area (a)
$A=(4+\pi)/2$, (b)-(d) $A=4+\pi$, lead widths (a) $d=0.125$, (b)-(d) $d=0.25$,
and entrance lead tilt angle in (d) $\delta=11.8^\circ$.}
\end{figure}

Over the past years transport experiments on phase coherent
mesoscopic systems have attained unprecedented levels of
sophistication. \cite{Revdot1,Revdot2,Revdot3} It is now possible
to precisely design and control the geometries of quantum dots
while reducing the effects of impurity scattering to a level where
the transport can be considered as purely ballistic (with scattering
only off the confining boundaries). \cite{Revdot1,Revdot2,Revdot3,Berry,Lee} For geometries which give rise to 
chaotic classical motion universal system properties are expected. 
\cite{Revdot2,Revdot3} 
Prototypical chaotic cavities (such as the stadium and the Sinai
billiard) do however 
feature geometrical symmetries that leave signatures 
on the transport properties,  
\cite{reflectionsymmetry,moises,blocksymmetry,schanze}  
which we investigate in the present
communication. In particular, we consider
systems with a {\em lead-transposing} reflection symmetry which 
interchanges the source and drain leads that couple the dot to the
electronic reservoirs [see, e.g.,~Figs.\ \ref{fig:systems}(a) and (b)], 
and contrast them with systems that do not
possess such a symmetry [see, e.g.,~Figs.\ \ref{fig:systems}(c) and (d)].

Our investigation is
based on random-matrix theory (RMT) for the scattering matrix.  
\cite{beenakker_review,pier_book} In fact, 
the present RMT problem is long-standing, dating back to almost 10 years ago
when first theoretical and numerical results were presented in Refs. 
\onlinecite{reflectionsymmetry,blocksymmetry}, respectively. These earlier 
works identified the correct 
invariant measure of the scattering matrix and they studied several 
statistical properties of the conductance. 
Here we give the complete solution of this problem by deriving the joint
probability density of transmission eigenvalues, which 
determine all stationary transport properties, for an arbitrary number 
of open channels supported by the leads attached to the quantum dot. 
The density of the transmission eigenvalues turns
out to take a particularly simple form. 
A striking 
signature of the solution is the absence
of any non trivial dependence of the ensemble-averaged
conductance and shot noise on the number of open transport channels. This 
prediction is confirmed numerically for
stadium billiards with different geometries. Our theory 
also explains
the deviations from standard random-matrix theory observed in earlier
numerical investigations. \cite{stefan}

The random-matrix theory of transport \cite{beenakker_review,pier_book} is
based on Landauer's scattering approach, which describes the transport
properties by the scattering matrix 
\begin{equation}
S=\left(%
\begin{array}{cc}
  r & t' \\
  t & r' \\
\end{array}%
\right),
\end{equation}
composed of amplitudes for transmission ($t$, $t'$) and
reflection ($r$, $r'$) between the channels in
the entrance and exit lead, respectively. We assume that both leads 
support the same number of open  channels $N$. 
We investigate the transport properties in terms of the 
eigenvalues $\tau_n$ of $t^\dagger t$ which determine the conductance
$G_N=\bar{G}\, \sum_{n=1}^N\tau_n$ (where $\bar{G}=2 e^2/h$ is the conductance
quantum) and the shot noise power \cite{buttiker_review} $P_N=(2e \bar{G} V)
\sum_{n=1}^N \tau_n(1-\tau_n)$ (where $V$ is the applied voltage).
For uncorrelated electrons the shot noise power is given
by the Poisson value $\bar{P}_N=2 e I $, where $I$ is the time averaged
current. Fermi statistics, however, induces electronic correlations and
deviations from the Poisson value which are customarily quantified by the Fano 
factor
\begin{equation}
\label{fanofactor} F_N = \frac{\langle P_N \rangle}{\langle \bar{P}_N
\rangle}=\frac{\langle \sum_{n=1}^N \tau_n(1-\tau_n)\rangle}{\langle
\sum_{n=1}^N \tau_n\rangle}\,,
\end{equation}
where $\langle\ldots\rangle$ stands for ensemble or energy average.

\section{Joint probability distribution of transmission eigenvalues}

Throughout this paper we assume the absence of magnetic fields and
spin-orbit scattering, hence, we consider time-reversal
symmetric systems without Kramers degeneracy. 

\subsection{Chaotic quantum dots without spatial symmetries}

As a benchmark for
our results we use the well-established random-matrix theory for
such systems in absence of any spatial symmetries 
\cite{beenakker_review,pier_book} (see the geometries in the bottom row of
Fig.~\ref{fig:systems}). The scattering matrix is then a member of
Dyson's circular orthogonal ensemble (COE), \cite{mehta} and the
joint probability distribution of the transmission eigenvalues
$\tau_n$ is given by
\begin{equation}
\label{poft_joint}
W(\{\tau\})=\prod_{n<m} |\tau_m-\tau_n |\prod_i \tau_i^{-1/2}.
\end{equation}
For a large number of channels $N\to\infty$,
 the density of transmission eigenvalues $\rho_N(\tau)=\left< \sum_n^N \delta
 (\tau -\tau_n) \right>$ approaches the bimodal
distribution \cite{harold,rodolfo} 
\begin{equation}
\label{rholarge}
 \rho_N(\tau)\approx \rho_\infty(\tau)=N
\left(\pi\sqrt{\tau(1-\tau)}\right)^{-1}.
\end{equation}
This gives an ensemble-averaged conductance 
$\left< G_N \right>=G_\infty+O(N^0)$, where
$G_\infty=N\bar{G}/2$, and a Fano factor $F_N=F_\infty+O(N^{-1})$, where
$F_\infty=1/4$. 

The bimodal distribution Eq.\ (\ref{rholarge}) is only valid
asymptotically for large $N$. The leading finite-$N$ correction for the
conductance is the
well-known weak-localization correction
$\left< G_N \right> -G_\infty=-\bar{G}/4
+O(N^{-1})$. \cite{beenakker_review}  Similar corrections exist also for 
other transport properties such as the Fano factor. Generally, they can be
related to deviations of
$\rho_N(\tau)$ from $\rho_\infty(\tau)$, and 
are most pronounced for a small number of channels.

For a single channel ($N=1$), the density of transmission eigenvalues is
given by \cite{harold,rodolfo}
\begin{equation}
\label{rho_N1} \rho_1(\tau)= \frac{1}{2 \sqrt \tau} ;
\end{equation}
hence the conductance and shot noise power is given 
by  $\left< G_1 \right>/\bar{G}=1/3$, $\left< P_1\right>/(2e\bar G V)= 2/15$, 
respectively. The Fano factor is then $F_1=2/5$.

For two channels, from Eq. (\ref{poft_joint}),  we find
\begin{equation}
\label{rho_N2}
\rho_2(\tau) = 4 \tau -3\sqrt \tau +\frac{1}{\sqrt \tau} .
\end{equation}
Thus $\left<G_2\right>/\bar{G}=4/5$,  
$\left<P_2\right>/(2e\bar G V)= 9/35$, and $F_2=9/28$. 
These
results show that for a small number of channels the corrections
to the large-$N$ asymptotics are clearly noticeable. 

\subsection{Chaotic quantum dots with spatial symmetries}

We now turn to systems with a spatial reflection symmetry which
interchanges the leads, such as the geometries of Fig.\
\ref{fig:systems} (a) and (b).  

For a left-right symmetric structure, Fig. \ref{fig:systems}a, the 
scattering matrix has the structure 
\begin{equation}
\label{s_symmetric}
S=\left(
\begin{array}{cc} r & t \\
t & r
\end{array}\right) ,
\end{equation}
where $r$ and $t$ are symmetric $N\times N$ matrices. Matrices of
this structure can be cast into a block diagonal form by using the
rotation matrix \cite{blocksymmetry}
\begin{equation}
R=\frac{1}{\sqrt 2}\left(
\begin{array}{cc} 1 & 1 \\
-1 & 1
\end{array}\right) ,
\end{equation}
{\it i.e.},
\begin{equation}
\tilde S=RSR^T=\left(\begin{array}{cc} s_1 & 0 \\
0 & s_2
\end{array}\right) ,
\end{equation}
where $s_1=r+t$ and $s_2=r-t$ are symmetric unitary matrices. In
terms of these matrices, the transmission matrix is given by $t=
\frac{1}{2}(s_1-s_2)$ and the reflection matrix is given by $r=
\frac{1}{2}(s_1+s_2)$. 

The transmission eigenvalues are obtained
from the matrix
\begin{equation} \label{ttdagger} tt^\dagger =
\frac{1}{4}\left[2-s_1 s_2^\dagger- (s_1 s_2^\dagger)^\dagger
\right] ,
\end{equation}
which involves the unitary matrix $Q=s_1s_2^\dagger$. Note that $Q$ and
$Q^\dagger$ can be diagonalized
simultaneously. Hence, the
eigenvalues $\exp(i\theta_n)$ of $Q$ determine the transmission eigenvalues by 
$\tau_n=\sin^2 (\theta_n/2)$.

In random-matrix theory the matrices $s_1$ and $s_2$ are assumed
to be independent members of the COE. 
We then can show that the joint
probability distribution of the eigenvalues $\exp(i\theta_n)$ is
also given by the COE. \cite{karol_1,karol_2} 
The matrix $Q$ can
be symmetrized by the unitary transformation $Q'=s_2^{-1/2}Q
s_2^{1/2}=s_2^{-1/2} s_1s_2^{-1/2}$, which leaves the
eigenvalues invariant. Moreover, the COE is invariant under the
automorphism $Q' \to U^T Q' U$, where $U$ is an arbitrary unitary
matrix which we identify with $U=s_2^{-1/2}$. Hence the
eigenvalues of $Q$ inherit the COE statistics of $s_1$, and the
joint distribution function of transmission eigenvalues takes the
form 
\begin{eqnarray}
 &&W(\{\tau\})=\prod_k\left\{ [\tau_k(1-\tau_k)]^{-1/2}
\sum_{\sigma_k=\pm 1}\right\}
 \nonumber \\ && \quad{}\times 
\prod_{n<m}
\left|\sigma_m\sqrt{\tau_m(1-\tau_n)}-\sigma_n\sqrt{\tau_n(1-\tau_m)}\right|
. \nonumber \\
\end{eqnarray}

The probability density $\rho_N(\tau)$ can be obtained directly
from the uniform distribution of eigenphases $\theta_n$ in the
COE. This yields $\rho_N(\tau)=\rho_\infty(\tau)$ [given in Eq.\
(\ref{rholarge})] {\em exactly, for any value of} $N$. 
Hence the ensemble-averaged
conductance
is given by $\left< G_N\right>=G_\infty=N\bar{G}/2$, i.e., the 
weak-localization 
correction is absent. Moreover the Fano factor is given by
$F_N=F_\infty=1/4$. The absence of any $N$-dependent corrections is
in striking contrast to the previously discussed case of
asymmetric systems.

\subsection{Chaotic quantum dots with fourfold symmetries}

The results above also apply to systems with a 180$^\circ$ {\em rotational} 
symmetry mapping the leads onto each other. They can also be extended to 
incorporate further spatial symmetries which may be
present in addition to the lead-transposing symmetry. 
For systems with two leads, the only remaining case
is the fourfold symmetry as indicated in Fig.\
\ref{fig:systems}b. In this case, the $S$ matrix can be written as  
\begin{equation}
S=\left(
\begin{array}{cc}
S_e &  0 \\
0   &  S_o
\end{array}
\right),
\end{equation}
where $S_e$ and $S_o$ are the scattering matrices associated with
the even and odd channels, respectively. Each matrix $S_{e (o)}$
has the structure of Eq. (\ref{s_symmetric}); thus each of them  
can be analyzed by the same procedure as in the previous 
lead-transposing symmetry case and the statistical transport properties 
are given by the superposition of the even and odd subsystems. For a cavity
with a fourfold symmetry we therefore have $\left< G_N\right>=\bar G N/2$
and a constant Fano factor $F_N=1/4$.

\section{Numerical simulations}

We now compare our theoretical results to numerical simulations of
stadium billiards which feature the four different setups depicted in
Fig.\ \ref{fig:systems}.  The numerical data is obtained by solving the
Schr\"odinger equation using a modular recursive Green's function
method which allows for an efficient calculation of the scattering matrix
even for a large number of open channels. \cite{stefan_green}

The most drastic effects of the
lead-transposing reflection symmetry are expected for the case of a
single channel in the leads ($N=1$). Figure \ref{fig:systems} compares the
numerical probability distribution $\rho_1(\tau)$ with the analytical
predictions for the systems with and without 
a lead-transposing reflection symmetry (see upper
and lower panel, respectively). Note that for each of the two pairs 
of geometries the numerical results (i) show a clear signature of the
absence/presence of the symmetry and (ii) we can see a good agreement with 
the analytical predictions.

\begin{figure}
\begin{center}
\includegraphics[width=0.8\columnwidth]{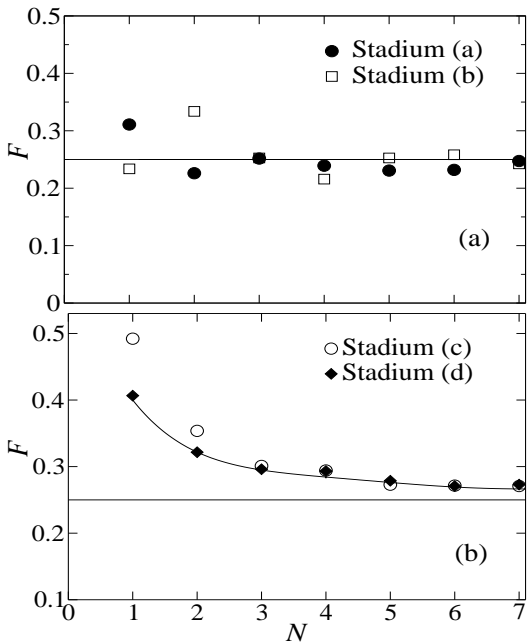}
\end{center}
\caption{\label{fano_180} Energy-averaged Fano factor in different mode
  intervals. Upper panel: Numerical data for systems with a lead-transposing
symmetry (see Fig.~\ref{fig:systems}a,b) compared with the
corresponding mode-independent random-matrix prediction ($F=1/4$, solid line). 
Lower panel: Analogous plot for systems without global
symmetry (see Fig.~\ref{fig:systems}c,d). The deviations to the RMT-result
  (solid line, numerically obtained from Eq. (\ref{poft_joint}) for $N>2$)
for the geometry depicted in Fig.~\ref{fig:systems}c can be
  explained in terms of the partial symmetry of short trajectories in the
  structure. }
\end{figure}

Several earlier studies have demonstrated that dynamical
signatures of geometries can also be identified in terms of the Fano
factor.  
\cite{eugene,marconcini,oberholzer,stefan,sim,henning,silvestrov,twor,jacquod,condmat}
Figure \ref{fano_180}a shows the Fano factor for the symmetric 
geometries as a function of the number of channels. For
the two systems with lead-transposing reflection symmetry our numerical
simulations show an overall constant behavior of the  Fano factor with the
number of channels, around $F=1/4$, in agreement with our 
modified random-matrix theory. 
The modes of the left-right symmetric cavity in Fig. 1(a) are
identical to the modes with even index of the cavity in Fig. 1(b). Therefore, 
the presence of the odd channels in our fourfold symmetric cavity does 
not change the
statistical properties of the transmission eigenvalues as predicted above.
The flat behavior of the Fano factor in the symmetric cavities is nicely 
contrasted by
the results for the two geometries without symmetry, Fig. \ref{fano_180}b: As 
predicted by
the conventional random-matrix theory (absence of spatial symmetries) 
the Fano factor decreases
as $N$ increases and approaches its universal value 1/4 for large
$N$. \cite{FN}

Note, however, that in Fig.~\ref{fano_180}b the Fano factor of the geometry
with the horizontally attached lead (Fig.~\ref{fig:systems}c) 
consistently lies above the RMT prediction for small channel numbers. This 
feature can be attributed to
the fact that in this geometry a significant fraction 
of back-reflected classical
tractories come in pairs of two equally long paths which are related to each
other by an up-down mirror reflection. This feature does, however, not apply to
those trajectories the mirror images of which are cut short by the exit lead
(see the two trajectory pairs in the inset of  
Fig.~\ref{fig:systems}c). Since this 
``partial'' symmetry enhances the
constructive interference of reflected paths but spares all transmitted
trajectories, the weak localization contribution to the conductance
and the shot noise power are increased. By tilting the entrance lead slightly 
(as in Fig.~\ref{fig:systems}d) this effect should be destroyed, resulting in
transport statistics which follow the RMT-prediction. Checking with 
 our numerical data in Fig.~\ref{fano_180} we note that this assessment
is indeed nicely corroborated. Given this excellent agreement, one might 
wonder why recently proposed corrections to RMT due to ``noiseless
scattering channels''
\cite{agam, silvestrov,twor,jacquod} do not apply in the present cases.
Due to the small ratio of lead width $d$ to cavity area $A$ in our geometries
(see caption Fig.~\ref{fig:systems}) the criterion for the emergence of
such fully transmitted or reflected channels, $N\gtrsim(k_F\sqrt{A})^{1/2}$,
\cite{silvestrov} is, however, not fulfilled in the energy regimes 
studied above (for the
highest energy considered here: $N=7$ and $(k_F\sqrt{A})^{1/2}\approx 15$). 

\section{Summary}

We have studied analytically and numerically the
effects of spatial symmetries on electronic transport properties
of ballistic lateral quantum dots, modeled by a quantum chaotic
cavity. Especially, we considered geometries with a symmetry which
maps the two leads onto each other. For such systems, random-matrix theory 
can be solved exactly for an arbitrary number of
channels $N$ in each lead. We predict that finite-$N$
corrections are absent for all transport quantities, such as the
conductance and the shot noise. This is confirmed by our numerical
simulations. 
 We further explored the effects of partial symmetries on 
transport which, as we showed, can yield significant corrections to 
the random-matrix predictions.

\section{acknowledgments}The authors gratefully
acknowledge helpful discussions with F.~Aigner, C.~W.~J. Beenakker, 
J.~Burgd\"orfer, and P.~A.~Mello. This work was supported by the European
Commission, Marie Curie Excellence Grant MEXT-CT-2005-023778
(Nanoelectrophotonics) and by the Ministerio de Educaci\'on y Ciencia, 
Spain, through the Ram\'on y Cajal Program. 

\end{document}